\documentclass[letterpaper,twoside,twocolumn,english,prl]{revtex4-1}
\usepackage[T1]{fontenc}
\usepackage[latin9]{inputenc}
\usepackage{xcolor}
\usepackage{amsmath}
\usepackage{graphicx}
\usepackage{esint}
\PassOptionsToPackage{normalem}{ulem}
\usepackage{ulem}


\usepackage{babel}
\begin{document}

\title{Mesoscopic  real space structures in spin-glass aging:  the Edwards-Anderson model.}

\author{Paolo Sibani$^{1}$ and Stefan Boettcher$^{2}$}

\affiliation{$^{1}$FKF, University of Southern Denmark, Campusvej 55, DK5230,
Odense M, Denmark~\\
 $^{2}$Department of Physics, Emory University, Atlanta, GA 30322,
USA. }
\begin{abstract}
Isothermal simulational  data for  the 3D Edwards-Anderson spin glass  are collected at 
several  temperatures below $T_{\rm c}$ and,
in analogy with a recent model of dense colloidal suspensions,
 interpreted  in terms of clusters of contiguous spins overturned by quakes,
non-equilibrium events linked to record sized energy fluctuations.
We show numerically that, to a good approximation, these quakes are
statistically independent and constitute a Poisson process whose average grows logarithmically 
in time. The overturned clusters  are local projections on one of the two ground states of
the model, and 
 grow likewise logarithmically in time.  Data collected at different temperatures $T$ can be collapsed 
  by scaling them with $T^{-1.75}$, a hitherto unnoticed feature of the E-A model, which we 
  relate on the one hand  to the geometry of configuration space  and on the other to experimental memory and rejuvenation effects.
  The rate at which a cluster flips is shown to decrease exponentially with the size of the cluster, as 
  recently assumed in  a coarse grained model of dense colloidal dynamics.
  The evolving structure of  clusters   in real space
 is finally ssociated to the decay of the  thermo-remanent magnetization.
 Our analysis  provides an unconventional coarse-grained  description of  spin glass aging
as statistically subordinated to a Poisson quaking process and highlights 
  record dynamics as  a   viable   common  
   theoretical framework for aging in   different systems.
\end{abstract}
\maketitle

\section{Introduction\label{sec:Introduction}}

Intensely investigated in the last few decades, the multi-scale dynamical
process called \emph{aging} is widely observed in glassy systems subject
to a change of an external parameter, e.g. a thermal quench. While
 spin-glasses~\cite{Fisher88a,Nordblad97,Vincent07,Guchhait15}, colloidal
suspensions~\cite{Hunter12}, vortices in superconductors~\cite{Nicodemi01},
magnetic nanoparticles in a ferrofluid~\cite{Jonsson00} and ecosystems
\cite{Becker14,Andersen16} may have little in common in terms of
microscopic variables and interactions, strong similarities emerge 
in their aging phenomenology. For example, one point averages feature 
 a logarithmic time dependence~\cite{Amir12} which entails an asymptotically vanishing rate
 of change of   the corresponding observables and clarifies 
why aging systems deceptively appear in equilibrium for observation  times shorter
than their age. Secondly, two-time averages such as correlation and
response functions often possess an approximate dependence on the single
scaling variable $t/t_{{\rm w}}$~\cite{Sibani10}. Interestingly,
this property is shared  by the probability that a species is extant at times $t_{{\rm w}}$
and $t>t_{{\rm w}}$ in a model of  biological evolution~\cite{Andersen16}.

Thermal relaxation models associate the multi-scaled nature of aging
processes to a hierarchy of metastable components of configuration
space~\cite{Palmer84,Hoffmann88,Sibani89}, often described as nested  `valleys'
of an energy landscape. Local thermal equilibration is described in
terms of time dependent valley occupation probabilities~\cite{Sibani93},
which are controlled by transition rates over the available `passes'.
When applied to a hierarchical structure, such description gradually
coarsens over time as valleys of increasing size reach equilibrium.
That  barrier crossings are connected to 
record values in time series of sampled energies~\cite{Dall03,Boettcher05}
 is a central point in  record
dynamics (RD), a coarse-grained description  of aging which uses the statistics of  non-equilibrium events called \emph{quakes} to 
describe aging in different settings~\cite{Anderson04,Sibani06,Sibani13,Sibani13a}.

 In connection with spin-glasses, RD has predictions 
describing  Thermo-Remanent
Magnetization (TRM) data~\cite{Sibani06} and  explaining their
observed \emph{sub-aging} behavior~\cite{Sibani10}, i.e. their deviation
from $t/t_{{\rm w}}$ scaling. In this work we explicitly check its basic assumptions 
and use it 
to  provide a different perspective 
on an iconic model of glassy behavior,
 the Edwards-Anderson (EA) spin-glass~\cite{Edwards75}.

Usually more reliant on system specific details than their more abstract
configuration space counterparts, real-space models often build on
the properties of domains whose time dependent linear size $l(T,t)$
characterizes the aging process, see e.g.~\cite{Jonsson00,Berthier02}.
Independent of
the mechanism assumed for domain growth, degrees of freedom belonging
to the same domain are assumed to fluctuate around their thermal equilibrium
state, while those located in different domains have, for  a fixed time scale,  frozen relative
orientations. The functional form of $l(T,t)$ can be extracted from
simulational data using a four-point equilibrium correlation function~\cite{Berthier02}.

Specifically in the spin glass droplet model~\cite{Fisher88a}, domains
are defined in terms of projections onto 
the two available ground states. Since the time growth of $l(T,t)$
minimizes the free energy by decreasing the domain wall length, the
droplet model views domain growth in a spin glass  as homologous to 
the scale-free coarsening process of a ferromagnet
at its critical temperature.

 Note however that,
while the interior of a ferromagnetic domain only harbors
local excitations of the ground state, analyses of small short-ranged
spin glass systems~\cite{Sibani94} indicate that each domain accommodates
a multitude  of metastable configurations. The same conclusion can be reached from a 
more recent enumeration of all the metastable configurations of E-A models of different 
linear sizes~\cite{Schnabel18}.
It thus seems questionable
that domain walls provide the main contribution to free energy
barriers in a spin glass. Finally, the droplet model leaves no room
for the temporally intermittent and spatially heterogeneous events
now recognized as key features of glassy dynamics~\cite{Schweizer07}.

From data analyses,   real space length scales  in aging systems
 are linked to   the
equilibrium correlation length of their metastable states, and recent numerical~\cite{Janus09,Janus17}
and experimental~\cite{Guchhait17,Zhai17} efforts utilize correlation
and response functions to describe the growth of 
 correlated domains.
Inspired by a recent model of colloidal aging~\cite{Boettcher11,Becker14a},
we use a different approach to identify growing real space structures in the E-A spin glass and 
argue that these are the   coarsening 
 variables  controlling aging  by linking them to  
 TRM data.
 
In models of dense colloids~\cite{Boettcher11,Becker14a} clusters of
contiguous particles, which gradually grow by accretion and suddenly
collapse through quakes, fullfill this dynamical role, while  the microscopic particle motion 
is only described
statistically  through a size dependent cluster collapse rate. The
crucial assumption that this rate decreases exponentially with cluster
size, corresponding
to the likelihood of a spontaneous fluctuation of that size, reproduces
the available numerical and experimental evidence on dense hard sphere
colloids.
As well, pertinent RD predictions, including a logarithmic time growth of
the average cluster size,
are obtained. A recent re-analysis~\cite{Robe16} of experimental
evidence shows that the quaking rate in dense colloidal suspensions
decreases as $1/t$, which is the basic claim from which RD predictions
flow. The experimental evidence was confirmed with molecular dynamics simulations of such a colloid~\cite{Robe18}.

To buttress our hypothesis, we analyze, as anticipated,   the dynamics of 
the E-A spin-glass~\cite{Edwards75},
a model with quenched randomness  microscopically
very different from a dense colloid.
Its  very well
studied behavior  is usually associated with  two competing
 theoretical approaches~\cite{Fisher88a,Newman96,Contucci12}
which, 
in spite of their differences, share  conceptual roots in  the equilibrium statistical
mechanics of  critical phenomena.  A unified description of 
 aging phenomenology requires, we believe, 
a much stronger focus on the statistics
of the rare non-equilibrium events that drive the dynamics in the full range
of parameters, e.g. temperature or density, where aging is observed.

Our simulations show: \emph{i)} That the energy changes associated
to quakes stand out from the overwhelming majority of energy fluctuations.
\emph{ii)} That quakes are statistically uncorrelated and occur at
a rate which is constant in \emph{logarithmic time}, as predicted
by RD. \emph{iii)} That suitably defined clusters grow on average  in proportion
to $\ln t$. The last result concurs with the behavior observed in~\cite{Boettcher11,Becker14a}
for a model of colloids.
Provided that the cluster size distribution is sufficiently peaked
around its mean, it also supports the latter model hypothesis that clusters
are overturned at a rate exponentially decreasing with their size.
Last but not least, our analysis  provides an approximate   description of spin glass dynamics
in terms of flipping clusters 
 which is  more complete   than previously available 
and covers  the TRM decay behavior.

The rest of the paper is organized as follows: In Section~\ref{Model}
the E-A model definition is stated for  the reader's convenience. In Section~\ref{Method} we summarize the
theoretical concepts used in our data analysis. Our numerical results are presented
in Section \ref{Results} and a real space coarse grained description
of the E-A spin glass dynamics is given in  Section~\ref{Cluster_dynamics}. Finally, Section~\ref{Implications} highlights similarities 
between our observed $T$  scaling of  energy fluctuations   and  experimental memory and rejuvenation  properties of
spin glasses.
 Section~\ref{Conclusion} provides a summary and draws conclusions.
\section{Model}

\label{Model} We consider an Ising E-A spin glass~\cite{Edwards75}
placed on a cubic grid with linear size $L=20$ and periodic boundary
conditions. Each of the $2^{N}$ configurations is specified by the
value of $N=L^{3}$ dichotomic spins, and has, in zero magnetic field,
an energy given by 
\begin{equation}
H(\sigma_{1},\sigma_{2},\ldots\sigma_{N})=\frac{1}{2}\sum_{i=1}^{N}\sum_{j\in{\mathcal{N}}(i)}J_{ij}\sigma_{i}\sigma_{j},\label{En_def}
\end{equation}
where $\sigma_{i}=\pm1$ and where ${\mathcal{N}}(i)$ denotes the
six nearest neighbors of spin $i$. For $j<i$, the $J_{ij}$s are
drawn independently from a Gaussian distribution with zero average
and unit variance. Finally, $J_{ij}=J_{ji}$ and $J_{ii}=0$. All
parameters are treated as dimensionless. This model has a phase transition from a paramagnetic
to a spin-glass phase at critical temperature which in Ref.~\cite{Katzgraber06} is 
estimated to be  $T_{\rm c}=0.9508$. The same reference reviews the  
different $T_{\rm c}$ estimates found in the literature.

\section{Method of analysis}
\label{Method} Starting from a configuration previously
equilibrated at temperature $T_{0}=1.25$, the system is instantaneously
quenched at time $t=0$ down to $T<1$. The ensuing aging process
is then followed for five decades in time. For aging temperature $T=.3,.4,.5,.6,.7,.75$
and $.8$, $512$ independent simulations are carried out and special
events, the quakes, are extracted from the trajectories thus obtained.
After defining a detection criterion
(see below), we check that quake events are uncorrelated and Poisson
distributed with an average proportional to $\ln t$. We then identify
clusters of spins that move in unison during the quakes, and from
those construct the average cluster size, $S_{{\rm Cl}}(t)$, as a
function of time.

{  The Waiting Time Method~\cite{Dall01} (WTM), a kinetic
MC algorithm which performs single spin flips with no rejections,
is used in all simulations. 
Similarly to the more widely used  Metropolis algorithm and its more recent  variants, e.g.
 parallel tempering~\cite{Katzgraber06a}, the WTM fulfils the detailed balance condition, and is by design guaranteed
to eventually sample the equilibrium distribution of the problem at hand.
Its performance in exploring  the  EA energy landscape at low $T$ was compared  in Ref.~\cite{Boettcher05} to  that
of Extremal Optimization~\cite{Boettcher01a}. These  two very different 
methods extracted the same geometrical features from the landscape, e.g. that
a record high energy barrier must be scaled in order to find a lower value of the lowest energy
seen `so far', or `best so far energy' $E_{\rm bsf}$ to which we shall return.
Being  calculated 
along the  trajectories as 
 differences between the energy of the current state and the $E_{\rm bsf}$, the above barriers  differ
conceptually from the overlap barriers   investigated in Refs.~\cite{Berg00}, which describe displacement  fluctuations
 in  thermal equilibrium.

In a jammed system as an aging spin-glass,  Metropolis  executes a large number
of unsuccessful trials (and the acceptance rate drastically declines), which the WTM avoids by
rank-ordering the execution time of all possible moves and 
then executing the one with the lowest execution time. 
Specifically, flipping spin $i$ at energy cost $\delta_{i}$ is associated
to a waiting time $w_{i}$ and the intrinsic time variable $t$ (flipping time)  of the WTM is  a real positive
number which sums up, at any point of the simulation, the times spent
`waiting' for all previous flips.
Each waiting time is
drawn from an exponential distribution
with average 
\begin{equation}
\langle w_{i}\rangle=\exp(\frac{\delta_{i}}{2T}).
\label{avWT_def}
\end{equation}
Hence, as  long as its  local environment 
 remains unchanged,  the  thermal flips  of each spin are  a memoryless  Poisson
process with the above average. This  seems  a physically appealing
description of  systems with many coupled degrees of freedom and implies that,
when a spin is reversed,  only the waiting and flipping times
of that spin and   its neighbors need to be   recalculated, while  all
others  can stay put.

Both the WTM and the Metropolis algorithm   lack a physical time scale, 
and their ability to empirically describe aging processes depends
on the temporal scale invariance of such processes, combined with the fact that  both  
methods seek the pseudo-equilibrium states in which aging system dwell most of the time.
Once the Metropolis algorithm has had a chance to query every spin,
it flips a set of spins similar to that flipped by the WTM.
For times of the order of  a MC sweep or larger,
the two methods are equivalent  and our $t$  corresponds to the number of MC sweeps~\cite{Dall01}.

The sequence of flips is however clearly different,
since Metropolis chooses the `next' flip candidate at random, while each  choice of the WTM can be influenced by 
the last flip:
Equation~\eqref{avWT_def} implies that any negative `barrier' $\delta_{i}$
which arise after a move creates a locally unstable situation where the involved spins quickly  flip. This process can iteratively
generate a series of negative $\delta_{i}$ values  in a local  neighborhood, triggering 
 event cascades whose
  short duration allows one to 
time-stamp   quakes with high  resolution.
The latter feature  is important when assessing  the  temporal statistics
of the quakes.
Besides being computationally  inefficient at low $T$, a Metropolis 
algorithm would express `times'  as integer
number of sweeps, which is at variance with time being  a real
variable  in a Poisson process. In contrast, WTM readily resolves 
sub-sweep timescales. 
 }

For short time intervals and at
low temperatures, the WTM dwells in real space neighborhoods
of local energy minima, and the sampled energy changes   feature a previously
unnoticed temperature scaling which is
found 
in most of our figures and explained in Section~\ref{Explanation}
in terms of the distribution of single flip energy changes available
near local energy minima.

\subsection*{Clusters and domains}
\begin{figure}
\hfill{}\includegraphics[width=0.9\columnwidth]{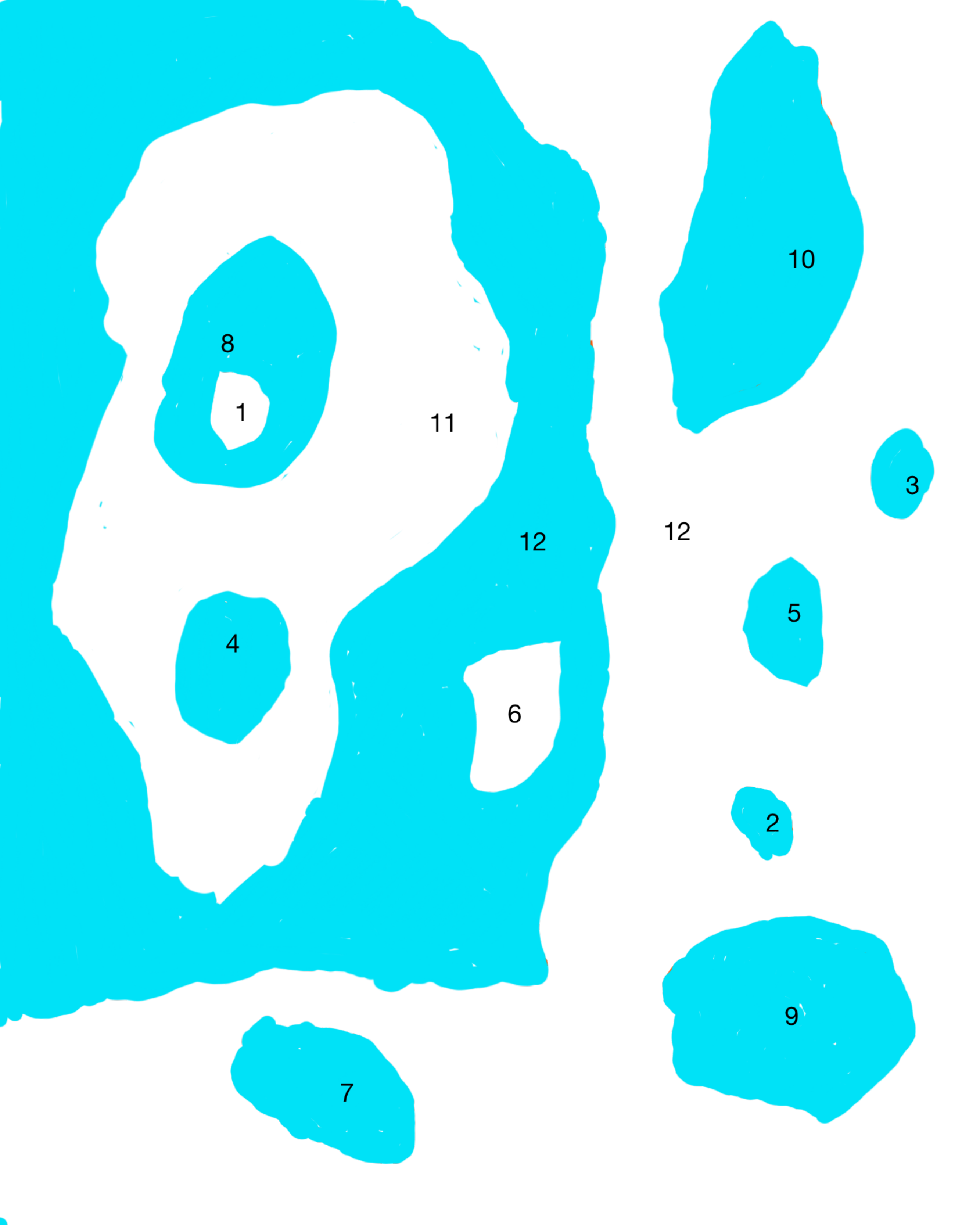}\hfill{}

\caption{\label{domains_fig} Depiction
of the  domain hierarchy in a hyper-plane of a 3d-Edwards-Anderson
spin glass during the aging process. Each numbered area represents
spin clusters with the same configuration as one of the two ground
states of the E-A spin glass. With the exception of area $12$, which
has two colors, each cluster is surrounded by a region of the opposite
color and takes up this color when overturned by a quake. In
this picture, randomly fluctuating, isolated spins have been suppressed.
A quake event amounts to filling in one of the inner-most domains
through flipping all its spins, thereby coarsening the otherwise self-similar
spatial hierarchy of domains-within-domains.}
\end{figure}

A local energy minimum configuration consists of  disjoint groups of
contiguous spins, our clusters, whose orientation is either the same
or the opposite as one of the two ground states, if one neglects, as we presently do, 
the  spins on the cluster  boundaries. Since each cluster
may contain sub-clusters of opposite orientation, a partially nested
structure is generated, reflecting the degree of hierarchical organization
of the system's configuration space~\cite{Sibani89,Sibani94}. The
situation is illustrated in Fig.~\ref{domains_fig}, using two dimensions
for graphical convenience. Excess energy relative to the ground state
stems from cluster interfaces and can be reduced in a thermally activated
process overturning gradually larger clusters. The free energy cost
of such reversals is mainly associated with barriers in the bulk of
each cluster, as we will explain below. In contrast, the cost of overturning a ferromagnetic
domain is mainly associated with the domain's interface.

Quickly reversible single spin flips similar to `in cage rattlings'
in a colloid are excluded from cluster configurations. Their long
term effects are subsumed into the statistics of the quakes which
provide the elementary moves, i.e. cluster flips, of the coarse-grained
dynamics we are about to describe. Since spins move together in a
quake, the final configurations of two successive quakes are compared,
all spins which changed orientation are identified and grouped into
clusters of spatially contiguous elements. Finally, clusters with
less than $5$ spins are discarded to minimize the risk of erroneously
counting reversible moves as part of a quake.

\subsection*{Quake detection protocol}
\label{detection}
Observation of non-equilibrium
phenomena is fundamentally tied to choosing the correct time and length scales.
This applies certainly to the aging process. On very large scales
macroscopic  variables seem to change in 
a  smooth and gradual manner. On 
intermediate scales aging systems appear in a state of quasi-equilibrium
punctuated by  increasingly
rare, intermittent quakes  that significantly (i.e., irreversibly)
relax the system and lead to overall structural changes.
The importance of these events 
 for the progression of the aging process was highlighted in \cite{Sibani05a}
 using a system-wide approach.
However,   since  quakes  unfold almost
instantaneously on an  intermediate time-scale, a more detailed 
investigation is needed to  explore the \emph{spatial} dynamic that facilitates
the quake. In the following we outline a protocol to zoom in more
closely into a narrower  time-window, as illustrated in Fig. \ref{fig:protocol},
where  the quake's footprint  is measured
from the difference  between the configuration it generates and that it 
inherits previous quake, see
Fig. \eqref{domains_fig}. This contrasts with equivalent aging experiments
on structural glasses such as colloids, where  spatial traces of quakes 
are faint.

Our method of data analysis identifies quakes on the fly from an evolving
trajectory and treats them, approximately, as instantaneous events.
The identification process involves a number of computational choices,
which are all based
on the following assumptions: Using $\ln(t)$ rather than $t$ as
independent variable transforms the quakes into a memoryless Poisson
process. Accordingly, successive quakes are statistically independent,
and if $t_{k}$ is the time of occurrence of the $k^{{\rm th}}$
quake, the `logarithmic waiting times' $\Delta{\rm ln}_{k}=\ln(t_{k})-\ln(t_{k-1})=\ln(t_{k}/t_{k-1})$
are independent stochastic variables with the same exponential distribution.
Correspondingly, the logarithmic rate of quakes is  constant.

In Ref.~\cite{Sibani05a,Sibani07} energy differences were sampled
over time intervals of duration $\delta t$, chosen much smaller than
the system age but larger than the decay time of the energy autocorrelation
function. {On this intermediate
time-scale, intermittent
events were distinguished from equilibrium fluctuations based on their
correspondence to rare, negative and numerically large energy changes
without resolving the quake event itself. In our case, we provide precise
values for the onset
times of quakes by explicitly connecting them 
to the extremal value of the `energy barrier' function discussed in
Refs.~\cite{Dall03,Boettcher05}. For that purpose, energy changes
in  close proximity
of local energy minima are monitored by choosing $\delta t$ now
much \emph{shorter} than the energy autocorrelation decay time, such
that neither equilibrium fluctuations nor quakes can unfold within
a single $\delta t$. Energy changes  measured within such a short
$\delta t$ without reference to barrier-height feature a perfect
normal distribution over many orders of magnitude, see Fig. \eqref{FL_stat}.
That  the width  of this distribution  scales anomalously with with temperature
confirms  that the energy changes sampled are not  equilibrium fluctuations.
\begin{figure}
\vspace{-5mm}
\hfill{}\includegraphics[bb=0bp 400bp 550bp 612bp,clip,width=1\columnwidth,height=0.4\columnwidth]{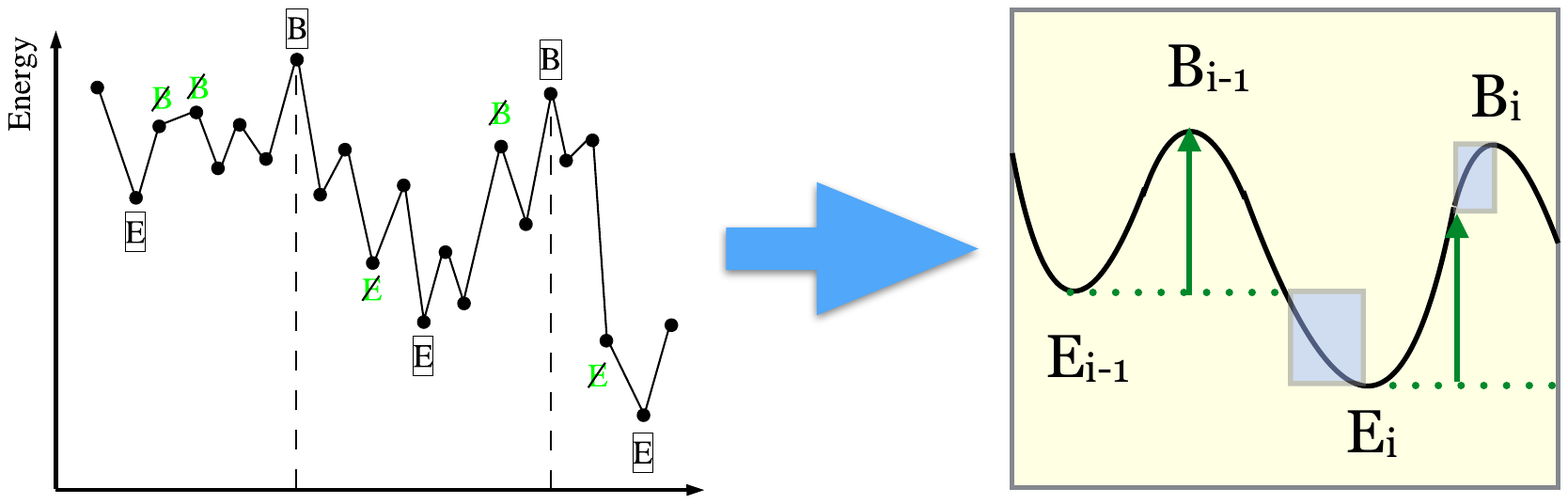}\hfill{}
\label{fig:CoarseValley}
\vspace{-7.5mm}
\caption{\label{fig:CoarseValleys}  The instantaneous energy $E(t)$ of the  system fluctuates widely while decaying slowly overall (left panel). The lowest energy $E_{{\rm bsf}}(t)=\min_{t}[E(t)]$, and the highest barrier $\max_{t}[E(t)-E_{{\rm bsf}}(t)] $
 ever seen  up to time $t$ are  marked by  $E$ and $B$, respectively. In Refs.~\cite{Dall03,Boettcher05}, intermediate records were stricken (crossed-out green letters) and the  last $B$-record before the next $E$, or the last $E$-record before the next $B$
 were kept to coarse-grain the states visited  into "valleys" entered and exited at barrier-crossings $B_{i-1}$ and $B_i$ and to demarcate the  catchment basin of the local minimum at $E_i$, as shown in the right panel. Here, we focus on the record-producing parts of the trajectory enclosed in the shaded boxes. In the lower box $E(t)$ begins to undercut the previous minimum, $E_{i-1}$, until $E_i$ is reached and in the 
 upper box  it exceeds the previous barrier record (up-arrow) until $B_i$ is reached.} 
\hfill{}\includegraphics[bb=0bp 120bp 650bp 612bp,clip,width=1\columnwidth, height=0.6\columnwidth]{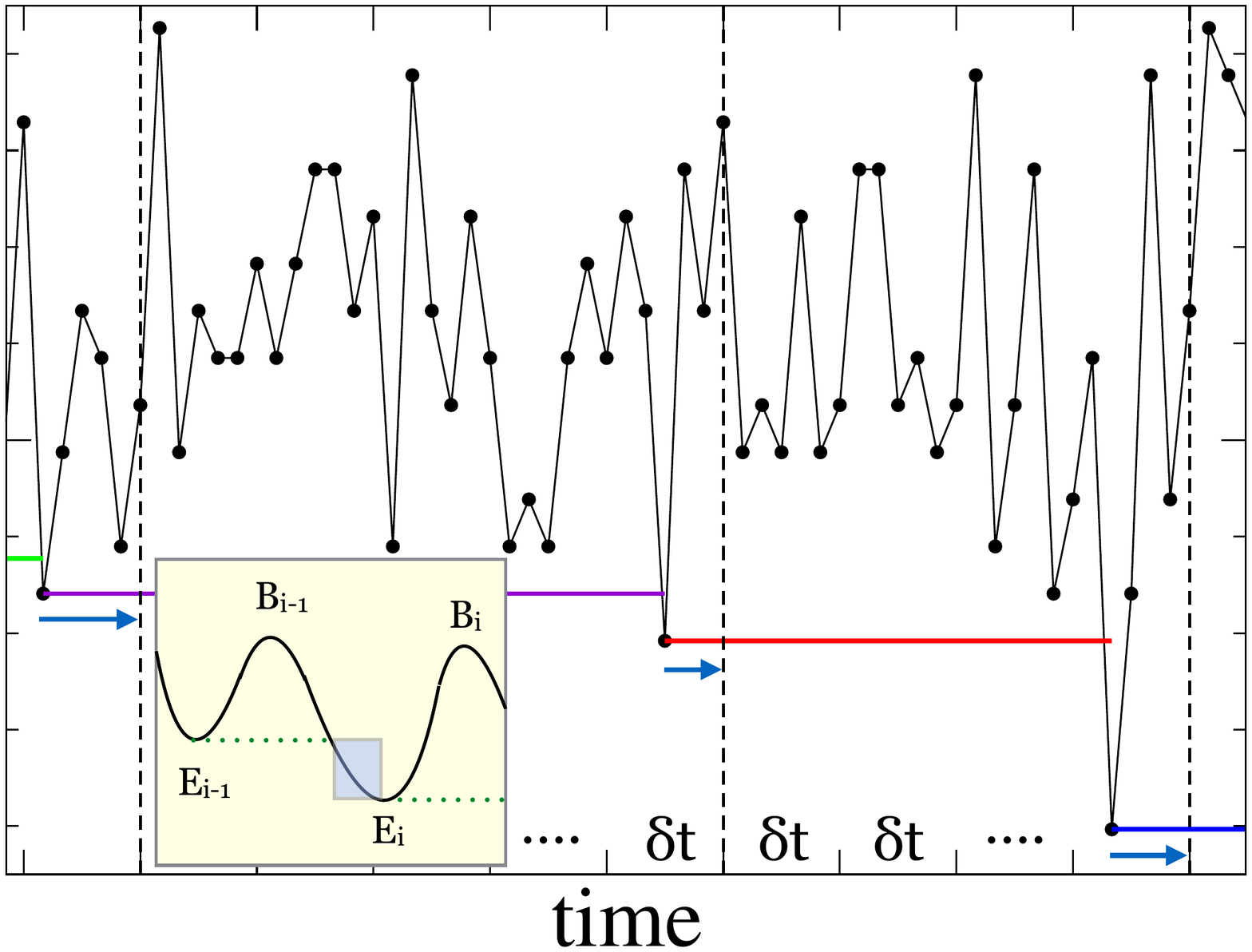}\hfill{}
\vspace{-2mm}
\hfill{}\includegraphics[bb=0bp 120bp 650bp 612bp,clip,width=1\columnwidth,height=0.6\columnwidth]{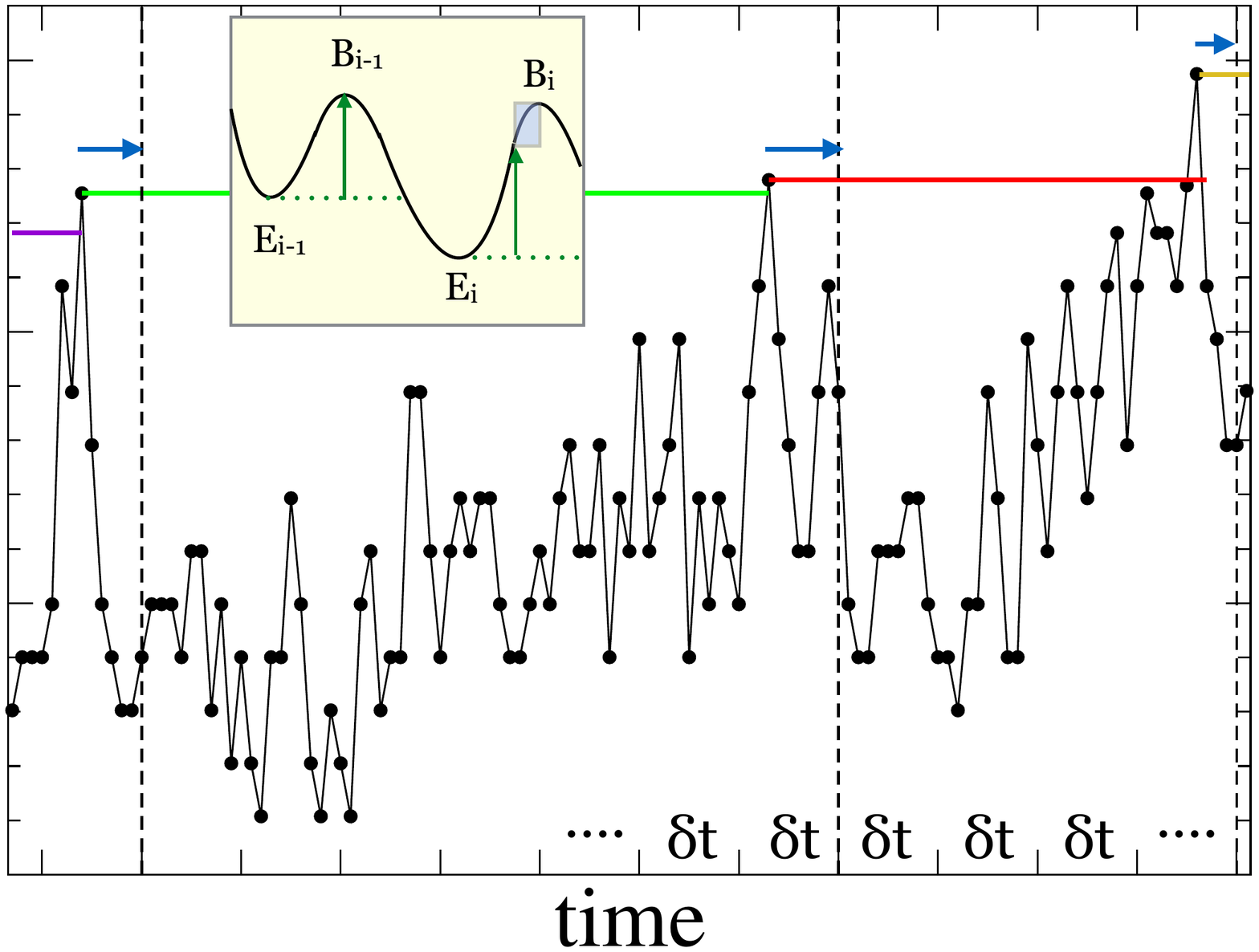}\hfill{}
\vspace{-5mm}
\caption{\label{fig:protocol} On-the-fly detection
of quakes while reaching new energy
minima $E_{i}$ (top panel) or  barrier records $B_{i}$ (bottom
panel). Within the respective ranges (shaded boxes in insets), a progression
of new records, either of $E_{{\rm bsf}}(t)$ (top) or of $b(t)$ (bottom),
is reached through quakes. In top (bottom) panel, once the energy
signal reaches below (above) the previous record, a quake event commences,
marked by a colored horizontal line. To capture the footprint of such
a quake, we record the spin configuration at the end of those time-intervals
$\delta t$ that contain a record (vertical dashed lines). The spin orientation changes
 between  consecutive quakes provide the spatial
extent of the intervening quake. The sub-interval duration  $\delta t$ used in 
the simulation is $\delta t=0.999$.}
\end{figure}

In contrast, to capture an actual quake, we have to use a specific
trigger, described in Figs.~\ref{fig:CoarseValleys} and \ref{fig:protocol}. Following Refs.~\cite{Dall03,Boettcher05},
we consider the barrier function $b(t)=E(t)-E_{{\rm bsf}}(t)$, where
$E_{{\rm bsf}}(t)=\min_{t}[E(t)]$ is the lowest energy ever seen
up to time $t$. According to Ref.~\cite{Dall03}, the entry and
exit times of a trajectory in and out of a valley in the energy landscape
can be evinced from the sequence of configurations where $b(t)$ and
$E_{{\rm bsf}}(t)$ reach their maxima and minima, respectively. 
As the description in Fig.~\ref{fig:CoarseValleys} demonstrates, the most recent barrier record $B_{i}$ only becomes recognized
as such when the next minimum is reached and, correspondingly, the
latest $E_{i}$ is certified as such only after $b(t)$ achieves a
new record. Thus, this classification scheme requires a-priori knowledge
of the entire time series of energy values, which we want to avoid.
Furthermore, we do not only focus on exit and entry points of valleys
in configuration space, but wish to identify the spatially localized
non-equilibrium events which provide the path approaching $E_{i}$
and $B_{i}$, respectively marked by a shaded box in the insets of
Fig.~\ref{fig:CoarseValleys}. 
Approaching $E_i$, $E(t)$ achieves a sequence of new
$E_{{\rm bsf}}(t)$ after the latest record barrier crossing. In turn, the function $b(t)$ reaches new records after
the latest minimum $E_{{\rm bsf}}(t)$ become fixed and $B_i$ is approached. Typical sequences
of $E(t)$ within those regimes are depicted in the main panels of
Fig. \ref{fig:protocol}. For either regime, we stipulate
that, if $E_{{\rm bsf}}(t)$ or $b(t)$ achieve a new record value
at $t=t_{r}$, a quake is unfolding. As soon as $t$ then reaches
the upper boundary of the sub-interval containing $t_{r}$, i.e.,
$t\leq t_{r}<t+\delta t$, that quake is deemed to have ended and
the system's configuration is saved. We then repeat this procedure
for the next record, until $E_{i}$ or $B_{i}$, respectively,  is reached
and continue the process in valley $i+1$ at later times. From the energy
differences $\delta E_{q}(i),\;i=1,2\ldots N$ between the current
and the previously saved configurations one easily finds the total
energy change connected to the quake and the positions of the participating
spins. The statistical error in the procedure comes from unrelated
spins which flip and participating spins which flip twice.

The above detection scheme allows a precise assessment of quake times,
and does not use  threshold values to discriminate quakes
from quasi-equilibrium thermal fluctuations. The arbitrary subdivision of the observation interval into
sub-intervals of length $\delta t$ 
determines when a quake ends, but has only a minor effect on the measured values of 
inter-quakes times, which are typically much longer than $\delta t$.
Finally, reaching the different energy records which define our quake detection technique
also requires tortuous paths, which are tantamount to entropic barriers. These are not
shown in Figs.~\ref{fig:CoarseValleys}-\ref{fig:protocol}, but are important for the dynamics,
as argued in Section~\ref{Explanation}.

To conclude, the WTM is ideally
suited for our  measurements. It produces equivalent physical results
to random sequential MC, yet, WTM focuses more efficiently on the
few active spins that drive the dynamics. By ranking degrees of freedom
by their time for change, it targets on exactly those spins connected
within a quake and is able to time-stamp quakes with high accuracy.

\section{Numerical results}
\label{Results} 
{  After the initial quench  $T=1.25 \rightarrow T<1$, the system 
is aged up to time $t_{\rm w}=100$ without taking any data.
 Data are taken in the interval  $[t_{\rm w},10^5] $ which is subdivided into $10^5$ 
 subintervals of duration  $\delta t=0.999$. This duration is an upper bound for  the temporal resolution of quake times, as explained in
 the `Quake detection protocol' section above.
As mentined,  $512$ independent simulations are carried out for statistical
reasons, all starting from the same equilibrium configuration. }

The first two  subsections below detail different types of simulational  results, and the last subsection
rationalizes the $T$ scaling form used to collapse all our data. All quantities specified below are dimensionless. 
\begin{center}
  \begin{tabular}{ |l| |  c| }
    \hline
    \multicolumn{2}{|c|}{Mathematical symbols used}\\ \hline
    $T, t$ & Temperature and time \\ \hline
    $\delta t$ & Short time interval  \\ \hline
    $\Delta$ & Energy change over  $\delta t$ \\ \hline
     $\Delta_{\rm q}$ & Quake induced energy change\\ \hline
     $\Delta{\rm ln}$ & Logarithmic waiting time\\ \hline
     $r_{\rm q}$ & Logarithmic quaking rate\\ \hline
     $R_{\rm q}(t)$ & Quaking rate $=r_{\rm q}/t$\\ \hline
      $r_{\rm cl}$ & Logarithmic rate of cluster growth\\ \hline
$n_{\rm q}(t)$ & Number of quakes up to time $t$\\ \hline
      $\mathbf{F}_{A}(x)$ & PDF of stochastic variable  $A$ \\ \hline
  \end{tabular}
\end{center}

\subsection{Energy fluctuations PDFs}
Energy fluctuations  sampled during isothermal aging at constant temperature $T$
have PDFs which change widely with $T$. As one would expect, the fluctuations are smaller 
the lower the temperature. Interestingly, their scaling is not linear in $T$, as would be the case
when dealing with equilibrium energy fluctuations, but involves instead  the  power law $T^\alpha$, where 
$\alpha=1.75$.
Let   $T^{-\alpha}\Delta$ denote the scaled
energy changes (per spin) sampled at temperature $T$ over an interval of a
very short duration $\delta t=0.999$. The length of this interval, which is much shorter
than those considered in \cite{Sibani05a} and far too short to straddle
equilibrium like energy fluctuations,
provides an upper bound for the duration of `instantaneous' quakes.

The seven estimated PDFs of 
$T^{-\alpha}\Delta$,
sampled at  seven  different aging  temperatures
$T=.3,.4,\ldots.7,.75$ and $.8$,  are
plotted in Fig.~\ref{FL_stat} using a light color (yellow)  and using, in order of increasing $T$,
squares, circles, diamonds, hexagrams, pentagrams, and down- and up-pointing
triangles as symbols. The dotted line is a fit of all these scaled PDFs to a Gaussian
of zero average. We note that the data collapse is excellent and that
the standard deviation of the Gaussian $\sigma_{G}\approx6.2\;10^{-3}$ is much smaller
than unity, the statistical spread  of the  coupling constants $J_{ij}$. This confirms 
that the sampled  energy changes  are
strongly constrained, as expected.

Quake induced energy changes $\Delta_{{\rm q}}$ occur over the 
 time intervals of varying length which 
 stretch from one quake to the next. Positive and negative values of
$\Delta_{\rm q}$ are associated with the system's energy increasing or
decreasing beyond its previous maximum or minimum, respectively. The
average effect of a quake is however an energy loss.  

The empirical PDFs of $T^{-\alpha}\Delta_{{\rm q}}$ are shown using
the same symbols as for the Gaussian changes, but a darker color (red).
For negative values of the abscissa these PDFs feature the exponential
decay  given by the fitted line, which  is reminiscent of
the intermittent tail seen in~\cite{Sibani05a}.
In this case, the  scaling with $T^{-\alpha}$ narrows but does not fully eliminate the spread of the
data. 
\begin{figure}
\hfill \includegraphics[bb=20bp 170bp 570bp 600bp,clip,width=1\columnwidth]{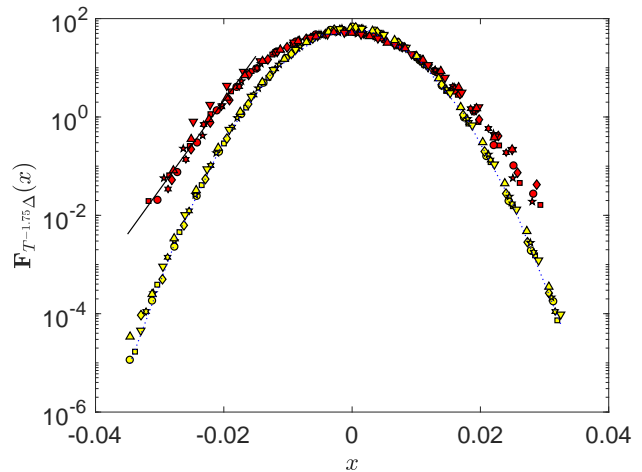}\hfill{}
\caption{Seven
PDFs of energy fluctuations  $\Delta$ collected at aging temperatures
$T=.3,.4,\ldots.7,.75$ and $.8$ are collapsed into a single Gaussian
PDF by the scaling $\Delta\rightarrow T^{-\alpha}\Delta,\;\alpha=1.75$, and plotted 
using  a logarithmic vertical scale. 
The data plotted with yellow symbols are fitted by the   Gaussian shown as a dotted line.
This  Gaussian  has  average $\mu_G=0$ and standard deviation $\sigma_{G}\approx6.2\;10^{-3}$.
Data plotted with red symbols  represent  quake induced energy fluctuations $\Delta_{{\rm q}}$ and, for negative values of the abscissa,
  have estimated probabilities close to the exponential PDF  shown  by the   
line.}
\label{FL_stat} 
\end{figure}
Isothermal aging was considered in~\cite{Dall03} for various spin-glass
models and the height of the energy barriers separating the neighboring
`valleys'  illustrated in Fig.~\ref{fig:protocol} was studied at different temperatures. Those data were collapsed
by $T^{1.8}$ scaling, a result which seems in reasonable agreement
with our present findings and is likely to have the same origin. 
\begin{figure}
\hfill{}\includegraphics[bb=20bp 180bp 560bp 600bp,clip,width=1\columnwidth]{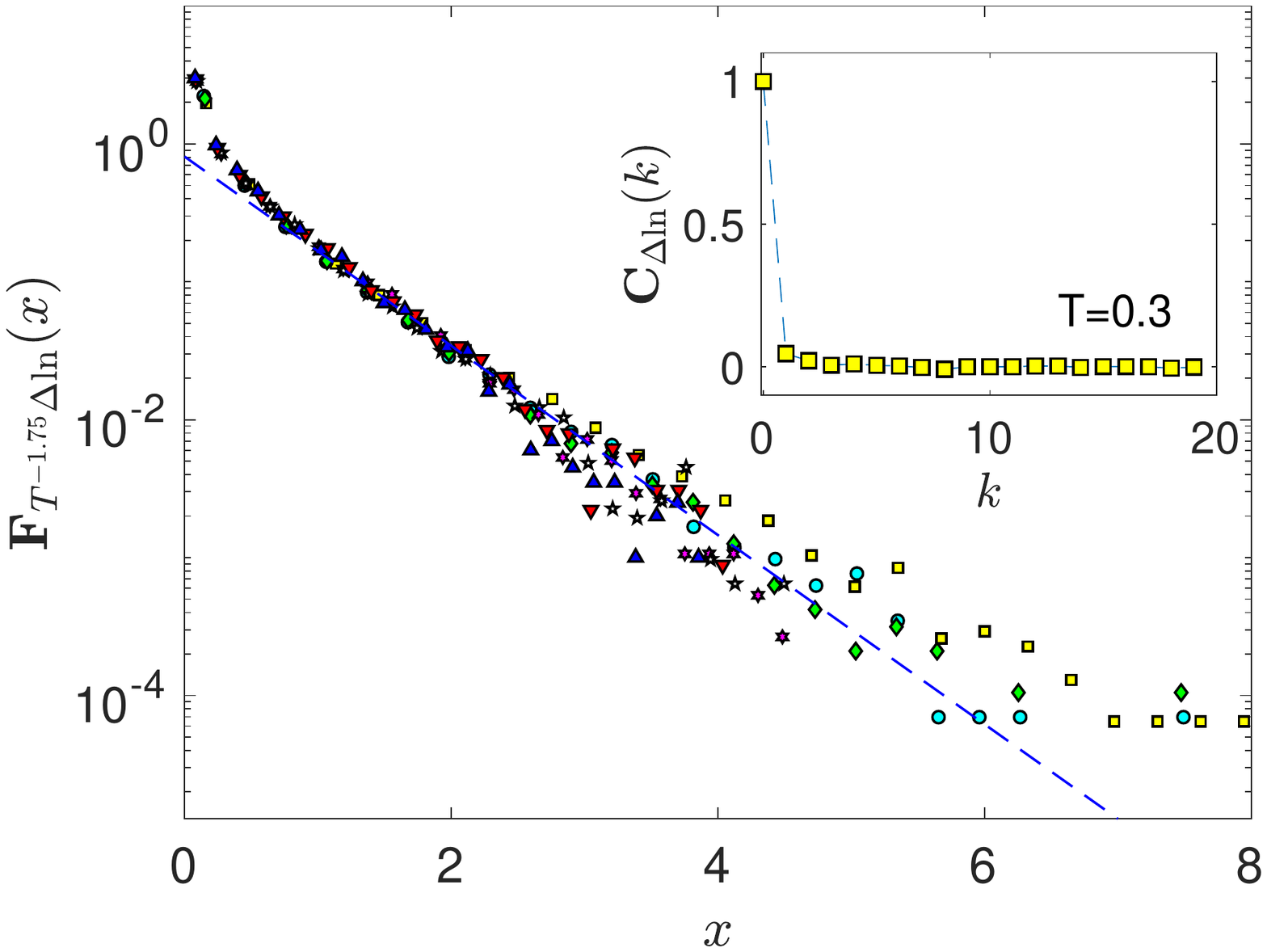}\hfill{}

\hfill{}\includegraphics[bb=20bp 180bp 560bp 600bp,clip,width=1\columnwidth]{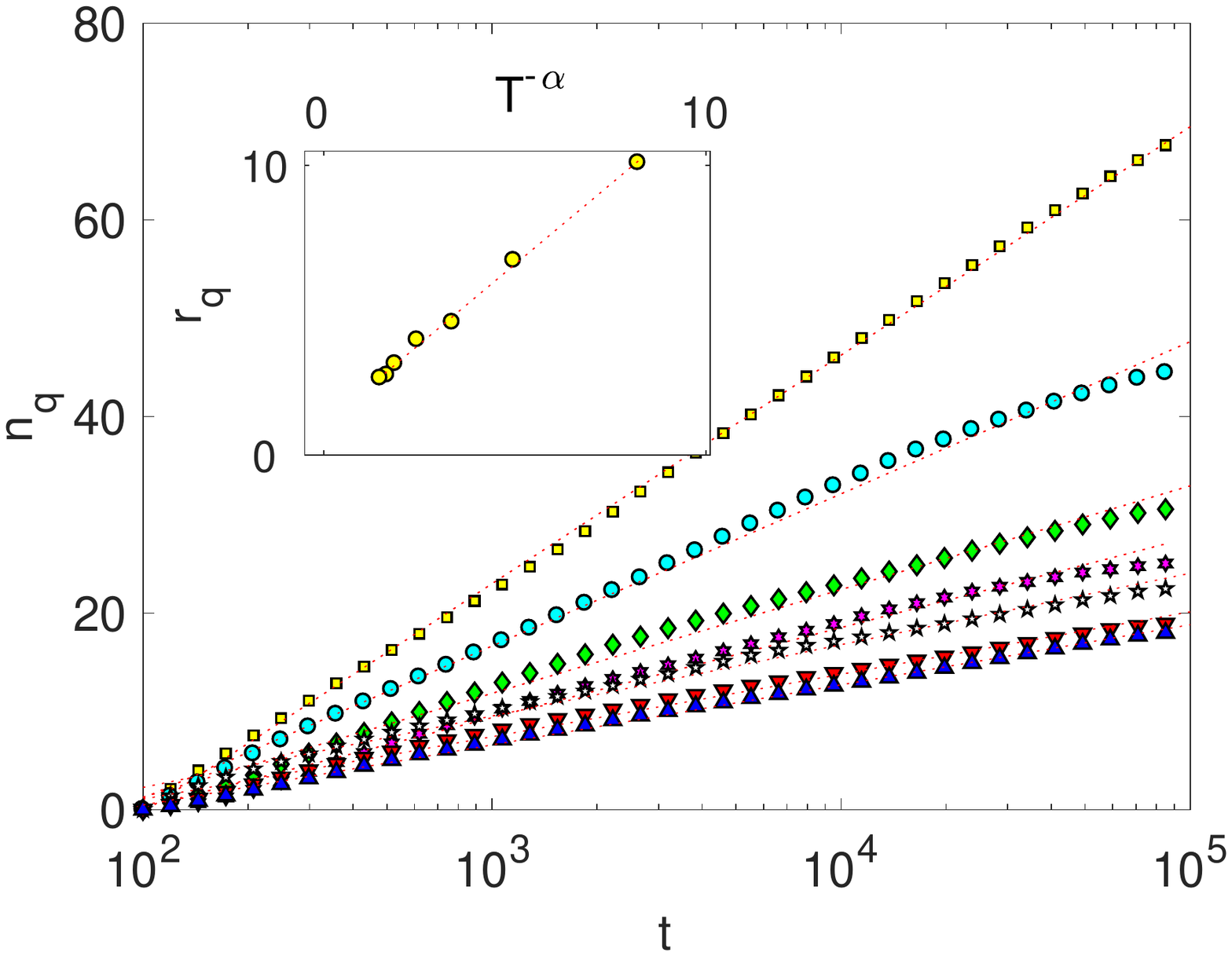}\hfill{}
\caption{Upper  panel. Symbols: PDF of scaled `logarithmic waiting times' $T^{-\alpha}\Delta{\rm ln}$, $\alpha=1.75$,
for the seven aging temperatures $T=.3,.4,\ldots.7,.75$ and $.8$.
Dotted line: fit to the exponential form $y(x)=.81e^{-1.57x}$. Insert:
the normalized autocorrelation function of the logarithmic waiting
times is very close to a Kronecker delta function $C_{\Delta{\rm ln}}(k)\approx\delta_{k,0}$.
The data shown are collected at $T=.3$, but similar behavior is observed
at the other investigated temperatures. Lower panel: the number
of quakes occurring up to time $t$ is plotted with a logarithmic
abscissa, for all $T$ values, with the steepest curve corresponding
to the lowest temperature. Insert: The quake rate, obtained as the
logarithmic slope of the curves shown in the main figure, is plotted
vs. $T^{-\alpha}$, where $\alpha=1.75$. The dotted line is a fit with slope $1.11$. }
\label{DLTS} 
\end{figure}

Consider now the times of 
occurrence $t'$ and $t$ of two successive quakes, $t>t'$, and
form the logarithmic time difference $\Delta{\rm ln}=\ln(t)-\ln(t')=\ln(t/t')>0$,
called for short, \emph{log waiting time}.
If quaking is a Poisson process in logarithmic time, the corresponding
PDF, $F_{{\Delta}{\rm ln}}(x)$ is given theoretically  by 
\begin{equation}
F_{\Delta{\rm ln}}(x)=r_{q}e^{-r_{q}x},\label{quaking_r}
\end{equation}
where $r_{q}$ is the constant logarithmic quaking rate. The applicability of equation~\eqref{quaking_r}
has already  been  tested in a number of different systems, including
spin-glasses~\cite{Sibani07}. 

The upper panel of Fig.~\ref{DLTS} shows the empirical PDFs of our logarithmic
waiting times, sampled at different temperatures and collapsed through
the scaling $\Delta{\rm ln}\rightarrow T^{-\alpha}\Delta{\rm ln}$.
The resulting PDF is fitted by the expression
 $F_{T^{-\alpha}\Delta{\rm ln}}(x)=.81e^{-1.57x}$, which 
covers two decades of decay. Its  mismatch with 
 the correctly normalized expression~\eqref{quaking_r}
stems from  the systematic deviations from  an  exponential
decay  visible for  small $x$ values. These deviations   arise in turn 
 from quakes which occur in rapid succession,
and  produce values $\ln(t_{k}/t_{k-1})\approx 0$. The effect, which is 
most pronounced  at early times in the simulation, roughly doubles the assessed
number of quakes, and correspondingly lowers the fitted pre-factor from $\approx 1.6$
to $\approx 0.8$. It furthermore produces 
non-zero correlation values in the series of log-waiting times
at $k=1$ and, to lesser extent, $k=2$.

Treating closely spaced quakes as parts of the same dynamical
event leads to the corrected number of quakes $n_{{\rm q}}(t)$ 
occurring up to time $t$ which is shown
in the bottom
panel of the figure for seven different aging temperatures.
 The steepest
curve corresponds to the lowest temperature. The red dotted lines
are linear fits of $n_{{\rm q}}(t)$ vs. $\ln t$, and the insert
shows that the logarithmic slope of the curves is well described by
the function $r_{{\rm q}}=1.11 T^{-1.75}$. We note that the logarithmic
quake rate as obtained from the  exponent (not the pre-factor) 
of the  fit $y(x)=.81e^{-1.57x}$  is $r_{{\rm q}}=1.57T^{-1.75}$.
The two procedures followed  to determine the quaking rate are  thus mathematically but not numerically
equivalent: in the time domain  they give the same $T^{-1.75}/t$ dependence of the quaking
rate, but with  two different pre-factors. The procedure using the PDF of
the logarithmic waiting times seems preferable, due to better statistics. 

Glossing over procedural difference, we write
$r_{{\rm q}}=cT^{-1.75}$ where $c$ is a constant, and note that in our RD description the
number of quakes occurring in the interval $[0,t)$ is then a Poisson process
with average $\mu_{N}(t)=cT^{-\alpha}\ln(t)$.
Qualitatively, we see that lowering the temperature decreases the log-waiting times
and correspondingly  increases the
quaking rate. The quakes involve, however, much smaller energy differences
at lower temperatures.  Considering that $T^{-\alpha}\gg T^{-1}$,
we see that the strongest dynamical constraints are not provided by
energetic barriers. As detailed later, they are entropic in nature
and stem from the dearth of available low energy states close to local
energy minima. Finally, our numerical evidence fully confirms  the idea 
that quaking is a Poisson process whose average is proportional to the logarithm of time.
In other words, the transformation $t \rightarrow \ln t$ renders the aging dynamics
(log) time homogeneous and permits a greatly simplified mathematical
description.

\subsection{Growth and decay of real space clusters}

The mean cluster sizes shown in Fig.~\ref{Av_cls} are calculated
as follows: Spins reversed by a quake are grouped into one or more
spatially disjoint sets, each comprising adjacent spins. Each set
is a cluster, and a first average cluster size $\overline{C_{j}}(t)$
is computed as the arithmetic mean of the sizes of all clusters generated
at time $t$ during the $j^{{\rm th}}$
simulation. 
 In a second step, our data are temporally coarse-grained
by placing 
logarithmically equidistant time points
$t_{1},t_{2}\ldots t_{n}$  within the chosen observation
interval, and by treating the quakes occurring in the same log-time bin  as 
simultaneous.
 The averaged cluster size 
$\tilde{S}_{{\rm cl}}(t_k)$ is then calculated as the arithmetic mean
of all the $\overline{C_{j}}(t)$s for which $t_{k-1}<t<t_{k+1}$.
This whole procedure is repeated for  different
values of the aging temperature $T$. It follows that $\tilde{S}_{{\rm cl}}(t_k)$
is the average cluster size, conditional to a quake
happening near $t_{k}$. Multiplying the result with the corresponding
probability $r_{{\rm q}}$ yields the (unconditional) average
cluster size ${S}_{{\rm cl}}(t_k)$.

Figure~\ref{Av_cls} shows that 
\begin{equation}
\begin{split}
\tilde{S}_{{\rm Cl}}(t)&=r_{\rm cl}(T) \ln t=
c'T^{2\alpha}\ln t\Rightarrow \\
{S}_{{\rm Cl}}(t)&=cc'T^{\alpha}\ln t, \label{MainR}
\end{split}
\end{equation}
where $c$ and $c'$ are positive constants.
 The  rate at which clusters
are overturned in real time, as opposed to logarithmic time, is
 $R_{{\rm q}}(t)=r_{\rm q}/t=cT^{-\alpha}/t$.
Inserting $t=\exp(\frac{S_{\rm Cl}T^{-\alpha}}{cc'}) $ from Eq.~\eqref{MainR}, we then obtain
\begin{equation}
R_{{\rm q}}(t)=cT^{-\alpha}\exp(-\frac{S_{{\rm Cl}}(t)T^{-\alpha}}{c'c}),\label{big}
\end{equation}
which provides the anticipated exponential relationship between the typical
cluster size and the rate at which clusters of that size are overturned.
Eq.~\eqref{big} does not prove that a specific cluster will be overturned
at a rate exponentially decreasing with its size, but is compatible
with that statement, if the spatial distribution of cluster sizes
is narrow.

\begin{figure}
\hfill{} \includegraphics[bb=20bp 180bp 560bp 600bp,clip,width=1\columnwidth]{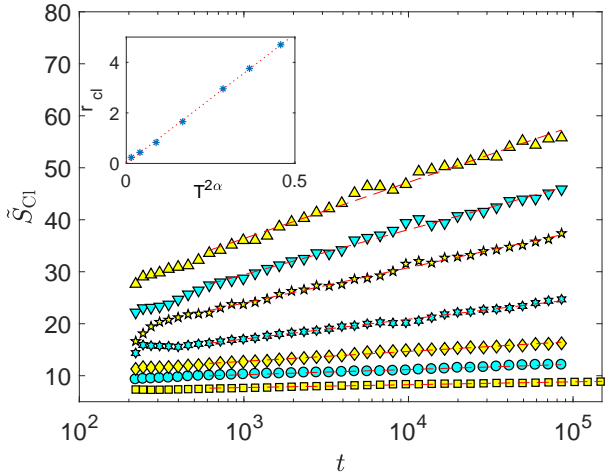}\hfill{}
\caption{Main plot: the average cluster size vs. the logarithm of time. The
data set, from bottom to top, are obtained at aging temperatures $T=.3,.4,.5,.6,.7,.75$
and $.8$. The red lines are linear fits of the data vs. $\ln t$.
The insert shows the slope of the linear fits vs. $T^{2\alpha},\;\alpha=1.75.$ }
\label{Av_cls} 
\end{figure}
\subsection{Origin of $T$ scaling}
\label{Explanation} 
To rationalize  the $T$ scaling of our data, we note that the  conditional waiting time $W|x$
for a spin to carry out a move with  energy change $x$ 
is exponentially  distributed  with average $e^{-\frac{x}{2T}}$, see Eq.~\eqref{avWT_def}, i.e.
\begin{equation}
p_{{\rm W|x}}(t)=e^{\frac{-x}{2T}}\exp(-t\;e^{\frac{-x}{2T}}).
\label{wtpdf0}
\end{equation}
The scaled energy changes $T^{-\alpha} \Delta$
shown in Fig.~\ref{FL_stat} have a 
 Gaussian distribution indicating  that $\Delta$  is a sum of several 
independent terms, all sampled over
short time spans of order one. Consequently, the positive energy changes selected must be of order  $x \approx T$,  and
the negative ones  are simply   their reversals. Let $g(x)$ be the probability density that 
an energy difference $x$
is associated to moves out of  a given configuration.
If the configuration is a local energy minimum,
very few `freewheeling' spins are present and,  for numerically small values of $x$,  $g (x)$ is zero for 
$x\le 0$ and increases with $x$  for $x>0$.
For configurations neighboring 
a local energy minimum, negative $x$ are available corresponding to moves back to the minimum
and the form of $g$ is reversed.
Glossing over the difference between local energy minima and their neighbors, we now assume that 
$g(x)\approx |x|^\beta$ for $\beta >0$ and, for  $x\propto T$,   find $\Delta \propto T^{1+\beta}$, which implies  that 
the $T$ dependence of the sampled energy differences 
can be scaled away by scaling  them with $T^{-\alpha}$, with $\alpha=\beta+1$.

Energy changes from one quake to the next are plotted in the same
figure, and have been similarly scaled. The $T^{-\alpha}$ scaling
does not fully collapse their PDFs as expected, since the time difference
between successive quakes is stochastic and typically much larger
than one. The result indicates however that a trajectory  triggering a quake
mainly consists of a  sequence of flips associated to  small
and reversible energy changes with the  `correct' $T$ scaling, rather
than  fewer but larger energy changes  associated to long waiting times.
In other words, entropic barriers play a large role in the dynamics.

Since, as we just argued,  the overwhelming
majority of the moves are associated with small time changes,
 the time between two quakes is a  sum of a varying, but large
number of short waiting times and inherits their   $T^{\alpha}$ dependence.
 The  number
of quakes preceding an arbitrary fixed time $t$ is then proportional
to $T^{-\alpha}$ as directly confirmed by the insert of the
lower panel of Fig.~\eqref{DLTS}, and indirectly by its  upper
panel, since the contents of the figures are mathematically equivalent.

\section{Spin clusters as dynamical variables}
\label{Cluster_dynamics} 
 The real space clusters  discussed  in the previous section are mesoscopic 
 objects which grow logarithmically in time.
In this  mainly  theoretical section, we use them as coarse-grained   variables,
and show that their dynamics explains 
the fit of TRM data 
 provided in~\cite{Sibani06}
as well as   other features of these macroscopic  data.
A table is included summarizing the notation used in this section.
\begin{center}
  \begin{tabular}{ |l| |  c| }
    \hline
    \multicolumn{2}{|c|}{Mathematical symbols in this section}\\ \hline
    $\lambda_i$ & $i$'th eigenvalue in corr. decay \\ \hline
    $w_i$ & weight of the corresponding term \\ \hline
    $r_{\rm q}(s)$&logarithmic rate of quakes hitting cl. of size $s$  \\ \hline
     $b$&logarithmic rate of quakes per spin \\ \hline
    $\kappa_s(t)$&no. of quakes hitting cl. of size $s$ in $[0,t)$  \\ \hline
     $p(s)$ & prob. that a cl. of size $s$ flips when hit \\ \hline
     $n_{\rm cl}(s,t)$ &no. of clusters of size $s$ present at time $t$\\ \hline
      $\mu_s(t_{\rm w},t)$ & average no. of hits to cl. of size  $s$ in $[t_{\rm w},t)$\\ \hline
      $\mu_s$ & same as above\\ \hline
  \end{tabular}
\end{center}

Adapting  Eq.(5) of ref.~~\cite{Sibani06}, TRM data  are described by the followin equation:
\begin{equation}
M_{{\rm TRM}}(t,t_{{\rm w}})=A_0\left(\frac{t}{t_{{\rm w}}}\right)^{\lambda_{0}(T)}\!\!\!\!\!\! +A_1\left(\frac{t}{t_{{\rm w}}}\right)^{\lambda_{1}(T)}\!\!\!\!\!\!+A_2\left(\frac{t}{t_{{\rm w}}}\right)^{\lambda_{2}(T)}\label{from_SRK0},
\end{equation}
where  the pre-factors $A_i$ and the  exponents $\lambda_i$ are  positive respectively   negative quantities.
Using that $\lambda_0$ is numerically very small, one further expands  the first power-law, obtaining 
\begin{equation}
M_{{\rm TRM}}(t,t_{{\rm w}})=A_0+a \ln (\frac{t}{t_{{\rm w}}} )+A_1\left(\frac{t}{t_{{\rm w}}}\right)^{\lambda_{1}(T)}\!\!\!\!\!\!+A_2\left(\frac{t}{t_{{\rm w}}}\right)^{\lambda_{2}(T)}\label{from_SRK},
\end{equation}
where $a=\lambda_0A_0\approx -1$ is independent of temperature in the available data range. Furthermore 
 $\lambda_{1}(T)$
and $\lambda_{2}(T)$ are weakly decreasing functions of $T$, with ranges close
to $-1$ and $-6$, respectively. 
Clearly, the logarithmic approximation to the first power-law eventually fails as $t/t_{\rm w}\rightarrow \infty$.
However, for the data range analyzed in~\cite{Sibani06} the logarithmic term is  dominant and the two remaining power-law terms
only provide fast decaying transients.

Since the gauge transformation $\sigma_{i}\rightarrow\sigma_{i}(t_{{\rm w}})\sigma_{i},\;J_{ij}\rightarrow\sigma_{i}(t_{{\rm w}})\sigma_{j}(t_{{\rm w}})J_{ij}$
maps the Thermoremanent Magnetization (TRM) into the correlation function
$C(t_{{\rm w}},t)=\sum_{i}\langle\sigma(t_{{\rm w}})\sigma(t)\rangle$,
modulo multiplicative constants, the two functions hold  for our purposes equivalent
information, and will be used interchangeably in the discussion.

Equation \eqref{from_SRK} was  justified in \cite{Sibani06} by the RD assumption that
aging  is log-time homogeneous and by then applying a standard eigenfunction expansion~\cite{VanKampen06} 
 for the magnetization autocorrelation function, alias TRM, namely
\begin{equation}
C(t,t_{{\rm w}})\propto\sum_{i}w(i)\exp(\lambda_{i}\ln(t/t_{{\rm w}}))=\sum_{i}w(i)  \left( \frac{t}{t_{\rm w}}\right)^{\lambda_{i}},
\label{anC}
\end{equation}
where $w_{i}\ge0$ and $\lambda_{i}\le0$. 
In view of the limited
accessible range of $\ln(t/t_{{\rm w}}))$, most modes in Eq.~\eqref{anC}
will either be frozen or have decayed to zero, leaving only a few active 
terms with an observable time dependence, 
precisely as assumed in \eqref{from_SRK}. 

The approach   leading to Eq.~\eqref{anC} implicitly describes 
the effects of the quakes by 
an unspecified  master equation, with  time replaced by its logarithm.
As a consequence, the exponential decays seen in many  relaxation processes 
are replaced by  power-laws, with no connection   to  a critical behavior.
Continuing along this line, we now construct the relevant  master equation and relate 
 its eigenvalues $\lambda_i$ to  real space 
  properties uncovered in our numerical investigation.
 Specifically, we  shall use  that  \emph{i)}  quakes are statistically independent events
 inducing  cluster flips, and that  \emph{ii)} they constitute a Poisson process. Since 
 spatial extensiveness then follows,  the rate of quakes hitting a sub-system, e.g. a cluster,
 is proportional  to the volume of the latter.
 
Some of the following  arguments rest on  unproven  hypotheses, i.e.
given that  a  quake hits  a cluster of size $s$, the latter is assumed to  flip with probability $p(s)$,
 a decreasing function of $s$, parametrised by 
\begin{equation}
p(s)=a_{0}+a_{1}s^{-1}+a_{2}s^{-2},
\label{flip_prob}
\end{equation}
where all three coefficients are positive. Further below, we argue that $a_0=a_1=0$.

Let   $\kappa_s(t)$ denote the number 
of quakes hitting a cluster of size $s$ and $n_{\rm cl}(s,t)$
the number of such  clusters  present at time $t$.
Finally,  $s_{\rm min}$ and $s_{\rm max}$ denote the sizes of the smallest and the largest 
clusters in the system.
The range of cluster sizes  is constrained by
the  condition $\sum_{s=s_{\rm min}}^{s_{\rm max}}s\;  n_{\rm cl}(s,t) =L^{3}$.
Finally, the total number  of quakes hitting the system between $t_{\rm w}$  and $t$
is $n_{\rm q}(t)=\sum_{s=s_{\rm min}}^{s_{\rm max}} \kappa_s(t)$.

Even though the   $\kappa_s(t)$  presumably share the $T^{-1.75}$ 
 temperature dependence  of  $n_{\rm q}(t)$,   the $T$ dependence of  $p(s)$ is unknown,
 as is that of the cluster distribution decay, which depends on  
  the products  $\kappa_s(t) p(s)$, see Eq.~\eqref{C_an2}.
 We therefore  gloss over $T$ dependences, but note that, in order to produce
 exponents  with a weak $T$ dependence~\cite{Sibani06},  $p(s)$ should increase with
 $T$ to counteract the strong decrease of the $\kappa_s(t)$. In other words, as the 
 temperature decreases the number of quakes increases but their  dynamical effect 
 is reduced.
 
As illustrated  in Fig.~\ref{domains_fig},
flipping a cluster, e.g. cluster 8, eliminates all
the sub-clusters  present in its interior,  in this case, cluster 1. 
To simplify our  treatment, this possibility is eliminated 
by assuming  that clusters are flipped
in order of increasing size. This is reasonable if, as we
shall argue, the logarithmic rate of cluster flipping decreases with
cluster size. Secondly, changes  in the size of a 
cluster induced by sub-clusters flipping in the cluster's interior 
are neglected.
The assumptions assign a dynamical significance to
the  hierarchy of cluster sizes present at $t=t_{\rm w}$ and allows 
 clusters of different sizes to develop independently  of each other.

Having neglected
the possibility that clusters flip in the `wrong' sequence,  a
cluster which flips contributes with its own size to the decay of the correlation
function. Furthermore, standard
arguments then  imply that the number $n_{\rm cl}(s,t)$ of clusters of size $s$ decays exponentially
in $\kappa_s(t)$. 
The correlation function and, equivalently,
the TRM,  are given by 
\begin{equation}
C(t_{\rm w},t)\propto 
\left\langle \sum_{s=s_{\rm min}}^{s_{\rm max} } s n_{\rm cl}(s,t_{\rm w})\exp(-p({s})\kappa_s(t))\right\rangle,
\label{C_an2}
\end{equation}
where the  constant ensuring the initial normalization  has been 
omitted and the average $\langle \ldots \rangle$ is performed over 
the distribution of each $\kappa_s(t)$.

The $\kappa_s(t)$ are  independent Poisson  variables  with expectation values 
\begin{equation}
\mu_{s}(t_{{\rm w}},t)=r_{{\rm q}}(s)\ln(t/t_{{\rm w}}),
\end{equation}
where $r_{{\rm q}}(s)$ is the logarithmic rate of quakes impinging
on a cluster of size $s$. The extensivity of  the quaking rates implies  $r_{{\rm q}}(s)=bs$
where $b$, a positive constant, is the logarithmic quake rate per spin.
As a consistency check, note that 
\begin{equation}
\sum_s r_{\rm q}(s)n_{\rm cl}(s,t)=b \sum_s s n_{\rm cl}(s,t)=b L^3 =r_{\rm q},
\end{equation}
the logarithmic quake rate for the whole system.

Each term of Eq.~\eqref{C_an2}  can be averaged independently  using 
\begin{equation}
\langle  \exp(-p({s})\kappa_s(t)) \rangle=e^{-\mu_{s}(t_{{\rm w}},t)} \sum_{j=0}^\infty \frac{\mu_{s}(t_{{\rm w}},t)^j}{j!}e^{-p(s)j},
\end{equation}
which evaluates to
\begin{equation}
\langle  \exp(-p({s})\kappa_s(t)) \rangle= \exp(-\mu_{s}(1-e^{-p(s)})).
\end{equation}
Expanding $e^{-p(s)}$ to first order, we finally obtain the 
contribution 
\begin{equation}
\langle  \exp(-p({s})\kappa_s(t)) \rangle \approx\exp(-\mu_{s}p(s))=\left(\frac{t}{t_{{\rm w}}}\right)^{-bsp(s)}
\end{equation}
to the average correlation function.

Summarizing,
\begin{equation}
C(t_{\rm w},t)\propto 
\sum_{s=s_{\rm min}}^{s_{\rm max} } s n_{\rm cl}(s,t_{\rm w})\left(\frac{t}{t_{{\rm w}}}\right)^{-bsp(s)},
\label{C_an3}
\end{equation}
which has  the same structure as Eq.~\eqref{anC}, with
the weight $w_i$ replaced by   the volume fraction $s n_{\rm cl}(s,t_{\rm w})$
occupied  by clusters of size $s$ at time $t_{\rm w}$
and  the eigenvalue $\lambda_i$ replaced by
$\lambda_s=-b s p(s)=r_{\rm q}(s) p(s)$, the flipping rate of clusters of size $s$.
 
Noting that Eq.~\eqref{flip_prob} entails 
 $\lambda_s=-b(a_0 s  -a_1-a_2 s^{-1})$,
we set $a_0=0$  on physical grounds, since 
the largest clusters would otherwise contribute to  the fastest 
decay of the correlation function.
The first non-zero term produces then a power-law decay term, $(t/t_{\rm w})^{-a_1 b}$,
while the next term gives a whole family of power laws with different
decay exponents, corresponding to the cluster size values 
initially represented in the system. 

To regain the form given in Eq.~\eqref{from_SRK0} we  set  $a_1=0$ and 
obtain  a sum of power-laws with exponents of decreasing magnitude
\begin{equation}
C(t_{\rm w},t)\propto  
\sum_{s=s_{\rm min}}^{s_{\rm max} } s n_{\rm cl}(s,t_{\rm w})
\left( \frac{t}{t_{\rm w}}\right)^{-a_2 b/s}.
\label{C_an3}
\end{equation}
Exponents corresponding to 
sufficiently  large clusters
will, to first  order in $ -a_2 s^{-1} \ln(t/t_{\rm w})$, all contribute to the constant and logarithmic terms
$ A_0+ a \ln(t/t_{\rm w})$ seen in Eq.~\eqref{from_SRK}. In summary, the
general form of the time dependence of the TRM data  given in Eq.~\eqref{from_SRK} is accounted
for by our qualitative arguments,  provided that a quake flips clusters of size $s$  with probability $p(s)=a_2 s^{-2}$.

The (mainly)  logarithmic decrease of the TRM data  is explained  using  our  EA model analysis  in terms of large clusters 
associated with  power-law terms with very small exponents, which can  be suitably expanded.
A different interpretation~\cite{Guchhait15} of the same  data uses the presence of   crystallites of different sizes
each size associated to an energy barrier and attributes the logarithmic decay of the TRM  to a wide
distribution of these barriers. Even though the E-A spin-glass lacks any crystallites,  the presence
of clusters of different sizes means that expanding the power-laws with small exponents  in  Eq.~\eqref{C_an3}  
yields, once the fast terms corresponding to small clusters have decayed,  
\begin{equation}
M(t_{\rm w},t)\propto  
A_0-  a \ln \left( \frac{t}{t_{\rm w}}\right),
\label{C_an4}
\end{equation}
where $a\propto  ( a_2 b)$.
This expression concurs with  the analysis of Ref.~\cite{Sibani06}, based on  the measurements of Ref.~\cite{Rodriguez03}
 if $a_2 b $ is independent  or nearly independent 
of $T$. Recalling that $b$ 
is the number of quakes per unit volume and per unit (log) time, an educated guess is $b \propto T^{-1.75}$,
in which case the probability that a cluster of size $s$ flips when hit by a quake should be $ p(s)=a_2/s \propto T^{1.75}/s$.
Note however that 
 the  $T$ dependence of the pre-factor of the 
logarithmic decay is  linear in Ref.\cite{Guchhait15}.

Most commonly denoted by $t$ in the literature, the `observation
time' elapsed after $t_{{\rm w}}$ is, in our notation, denoted by 
$t_{{\rm obs}}\stackrel{{\rm def}}{=}t-t_{{\rm w}}$. Interesting
geometrical features of the spin glass phase, such as the size of
correlated domains~\cite{Joh99,Janus17}, are associated to the `relaxation
rate' $S_{\rm R}(t_{{\rm obs}},t_{\rm w})$, defined as the derivative of the TRM with
respect to $\ln t_{{\rm obs}}$~\cite{Nordblad97}, and in particular to its broad maximum
at $t_{{\rm obs}}\approx t_{{\rm w}}$. To see the origin of the latter,
we derive the relaxation rate from Eq.~\eqref{anC} as 
\begin{equation}
S_{\rm R}(t_{{\rm obs}}/t_{{\rm w}})\propto\frac{t_{{\rm obs}}}{t_{{\rm w}}}
\sum_{s}|\lambda_{s}|w_{s}\left(\frac{t_{{\rm obs}}+t_{{\rm w}}}{t_{{\rm w}}}\right)^{\lambda_{s}-1},
\label{rel_rate}
\end{equation}
which is the product of an
increasing pre-factor $\frac{t_{{\rm obs}}}{t_{{\rm w}}}$ and a sum
of decreasing terms $\left(\frac{t_{{\rm obs}}+t_{{\rm w}}}{t_{{\rm w}}}\right)^{\lambda_{s}-1}$.
Each of these terms has a maximum at $t_{\rm obs}/t_{\rm w}=-1/\lambda_s$, and, together, they give rise to
 the broad maximum near $t=t_{\rm w}$ experimentally observed for the relaxation rate~\cite{Nordblad97}.

Using   $\lambda_s=- a_2 b/s$, 
and recalling that 
$w_s =s n_{\rm cl}(s,t_{\rm w})$, we   find that the relaxation
rate  for the value $t_{\rm obs}= 2 t_{\rm w}$ commonly used in the literature  is 
\begin{equation}
S_{\rm R}(2)\propto  \sum_{s=s_{\rm min}}^{s_{\rm max}} n_{\rm cl}(s,t_{\rm w})3^{-a_2 b/s} \propto 
\langle 3^{-a_2 b/s} \rangle,
\label{rr_max}
\end{equation} 
where the brackets denote an average over  the size distribution of clusters
present at $t=t_{\rm w}$.
Importantly, Eq.~\eqref{rel_rate} and\eqref{rr_max} show that 
the relaxation rate and its maximum both gauge  the characteristic size of the
clusters, or domains, present in the system at time $t_{\rm w}$. 
\section{Implications of $T^{1.75}$ scaling}
\label{Implications}
The $T^{1.75}$ dependence  of energy changes characterizing 
 isothermal trajectories  at different temperatures (see 
 Fig.~\ref{FL_stat}) implies  that the barriers separating the 
 parts of configuration space where these  trajectories unfold
are not easily surmounted by the 
 thermal ${\mathcal O}(T)$ fluctuations  available  in  quasi-equilibrium states.
 
To the best of the authors' knowledge, this anomalous scaling  has not been noticed
in other numerical simulations, except for a brief mention in~Ref.\cite{Dall03}, where a slightly different
scaling exponent was found.
However, as we argue below, the behavior  fits and partly explains the rejuvenation and memory effects   experimentally 
seen in spin-glasses~\cite{Jonason98,Mathieu10} under a change of temperature protocol.
  
 In~\cite{Jonason98}, the imaginary part of the magnetic susceptibility is measured
at  high frequency, $\omega > 1/t_{\rm w}$, while the system is cooled at constant rate through a range of low temperatures.
As such, this protocol  produces an  out of phase  (pseudo-)equilibrium magnetic susceptibility  $\chi ''(\omega,T)$,
which is utilized  as a reference or master curve. Importantly, the cooling process  is halted at temperature $T_1$ and
 the system is allowed to age isothermally 
for several hours, leading to a decrease, or `dip', of the susceptibility away from the master
curve. When  cooling is resumed, the measurements soon return to that curve,   a rejuvenation effect implying that states 
seen during the  aging process at $T_1$ have little influence  on those seen at other temperatures.  Furthermore,
a second aging stop at a lower temperature $T_2$  produces a second dip.
The striking memory behavior 
of the system is revealed when the system,  continuously  re-heated  without any aging stops,   re-traces the dips of the susceptibility
previously  created at $T_1$ and $T_2$.
  Similar rejuvenation and memory behavior is observed in  TRM traces~\cite{Mathieu10}.
These experiments
show  that  aging trajectories at different, not too close, temperatures are dynamically  disconnected.
Our  numerical data point, as anticipated,  in the same direction and offer at  the same time an explanation of the rejuvenation
part of the experimental findings.

\section{Summary \& Discussion}
\label{Conclusion} 
{  
 This work's  main  focus is to buttress    Record Dynamics (RD)~\cite{Anderson04,Sibani06,Sibani13,Sibani13a,Robe16}
as  a general method 
to coarse-grain  aging processes,
by analyzing numerical simulations from 
 a   model with quenched randomness. 
 Spin glasses are  iconic  systems,
 where a wealth of fascinating  phenomena illustrating   central aspects of complexity
have been experimentally uncovered 
 (see ~\cite{Nordblad97,Vincent07} and references therein), and  the
 E-A model was an obvious choice.

For  historical reasons,   traditional   interpretations
of both numerical and experimental  spin-glass data rely on adaptations of  
 equilibrium concepts, e.g. critical behavior and other properties of either~\cite{Vincent07} 
 the Parisi solution~\cite{Parisi83} 
of the  mean field Sherrington-Kirkpatrick model~\cite{Sherrington75},
or~\cite{Nordblad97}  the real space description of the E-A model~\cite{Edwards75}
proposed  by
Fisher and Huse~\cite{Fisher88a}.
Since RD relies  on the statistical properties of non-equilibrium events, the 
picture emerging from our investigations unsurprisingly  differs in some respects
from more established descriptions.

RD  tacitly assumes the existence of  a  hierarchy of free energy barriers
in configuration space~\cite{Sibani13a,Robe16} which, however,  
bears no direct relation  to mean-field spin-glass models and 
rests on  general arguments of dynamical nature~\cite{Simon62,Hoffmann88}, 
exemplified   by
a coarse-grained discrete  toy-model of `valleys within valleys', i.e.  thermal hopping  
on a tree structure~\cite{Sibani89}.

A connection between 
the ultrametrically organized pure states~\cite{Parisi83}
of the SK model, which are  intrinsically stable equilibrium objects, 
and  the metastable states
of  real spin glasses requires a degree of  funambulism. The needed tight-rope~\cite{Ginzburg86}
 is  provided in Ref.~\cite{Lederman91}, where  the spin-glass configuration space 
is depicted as a hierarchically organized set of metastable states.

 Treating an aging  spin glass as a critical ferromagnet in disguise 
is, we argued,  a dubious undertaking on two counts: \emph{i)} Even though the energy difference between two metastable states is 
associated to a domain wall,  the dynamical  barriers that hinder a reversal of the domain orientation
are not. They are instead associated to the 
interior of the domain. \emph{ii)} While the dynamics of a 3D spin glass looks critical
when $T_{\rm c}$ is approached from above,  once below $T_{\rm c}$  thermal equilibration is  chimeric and the physical   relevance  of 
the critical temperature  is  moot.

 Some descriptions, see e.g.~\cite{Vincent95}, model aging dynamics  as a 
  random walk  in a configuration space fraught with  traps 
 whose  exit times feature 
a  long tailed distribution~\cite{Bouchaud92} of unspecified origin. For a detailed 
discussion of continuous time random walks
and  `weak ergodicity breaking'  vs RD, we refer to~\cite{Sibani13}. Here we just note 
that RD traps all  have a finite depth, i.e. a finite average exit time,
but are typically visited  in order of increasing depth. Last but not least,  the quake, i.e. jump,  statistics  in RD is 
predicted from configuration space properties, rather than
simply assumed.

Keeping  our focus in mind, the experimental results discussed in some  detail~\cite{Sibani06,Guchhait15,Jonason98,Mathieu10} 
 are all directly connected to our findings. Secondly,
variants of the E-A model, e.g. binary coupling distributions, are  not discussed. Considering 
RD's broad applicability, it seems  plausible that  such models would yield qualitatively similar 
results. Some technical adjustments would however  be needed for  our definition of clusters,
as  the ground state is degenerated beyond 
 a  global inversion symmetry.

In a spin glass context, RD has been used to describe TRM experiments~\cite{Sibani06} and
numerical heat exchange data~\cite{Sibani05a}. In the present investigation,
quakes are operationally defined by  associating them  to record values of a
 suitably defined `energy barrier' function sampled during the simulations, as 
 graphically illustrated by Fig.~\ref{fig:protocol}.  That 
 these quakes are a Poisson process whose average grows with the logarithm of time
 is explicitly verified   in Fig.~\ref{DLTS}, which confirms the  basic assumption on 
 which RD relies.
 
 Neglecting easily reversed single spin excitations produces the coarse-grained picture
 we use, where
  every low temperature configuration 
 appears as   a collection of adjacent spin clusters, each 
 oriented as one of the two ground states of the E-A model. 
Clusters are identified from simulational data as groups of spins
which  change direction  during a quake  while 
keeping   their relative orientations unchanged.
 On average, the size of spin clusters overturned at time $t$ grows as $\ln t$ and the rate at which
a cluster is overturned decreases exponentially with its size.   This relation subsumes the effect of both 
entropy and energy barriers and establishes a  connection 
with our model of dense colloids~\cite{Boettcher11,Becker14a}.

A so far unnoticed property of the E-A model seen in Figs.~\ref{FL_stat}  and \ref{DLTS} 
  is that aging  data, e.g.  energy differences and logarithmic waiting
  times,   collected at different (low) temperatures   
 can be collapsed by scaling them with $T^{-1.75}$. 
 This property is explained with the form $g(x) \propto |x|^{3/4}$ which, for $x \approx 0$
 is assumed to describe the energy changes associated to moves to and from a local energy
 minimum configuration.
 In the simulations,  the WTM's
 dwells near  local energy minima, where it repeatedly samples this type
 of energy fluctuations.
  We argue that the dearth of available moves with a low associated energy change can
  explain the 
 rejuvenation part of memory and rejuvenation experiments~\cite{Jonason98,Mathieu10}: 
 Simply put, states explored during isotherrmal aging at different temperature are 
 separated by large dynamical barriers of entropic nature, and these 
 barriers are not easily overcome  by thermal equilibrium fluctuations, which scale linearly with $T$.
 
Finally, an approximate   real-space analytical description  
is developed using growing clusters as mesoscopic dynamical variables.
 Important elements
are that the logarithmic rate of quakes is an extensive and  time independent  quantity
and that, given that  a cluster is hit by a quake, it flips with a probability inversely proportional
to its size. Unlike the first assumption, the second  is only supported \emph{a posteriori} by the formula it produces,
which empirically
describes  TRM decay data~\cite{Sibani06}.
  
 Importantly, the power-law terms vanish fairly rapidly and the remaining logarithmic decay, which formally arises by expanding a possibly large group
of power-laws with small exponents, has a pre-factor which is  $T$ independent,
as in the experimental data analysis of~\cite{Sibani06} but in contrast with the formula given in~\cite{Guchhait15}.
A similar behavior~\cite{Nicodemi01,Oliveira05} is seen  in the temperature independence 
 of the magnetic creep rate of high $T_{\rm c}$ 
superconductors.

By focussing on non-equilibrium quakes and their statistics,
several  real space implications are brought forth of the  hierarchical
 energy landscape organization which RD relies on, and a clear relation emerges
 between configuration and real space pictures of spin-glass  dynamics, namely that
 increasingly scales in Hamming  and Euclidean distance become relevant as increasing dynamical barriers are 
 overcome.
}

\end{document}